# Quantitative Study on Exact Reconstruction Sampling Condition by Verifying Solution Uniqueness in Limited-view CT

Bin Yan, Wenkun Zhang, Lei Li, Hanming Zhang, Linyuan Wang[*]

*National Digital Switching System Engineering and Technological Research Center,*

*Zhengzhou, Henan, 450002, China*

**Abstract:** In limited-view computed tomography reconstruction, iterative image reconstruction with sparsity-exploiting methods, such as total variation (TV) minimization, inspired by compressive sensing, potentially claims large reductions in sampling requirements. However, a quantitative notion of this claim is non-trivial because of the ill-defined reduction in sampling achieved by the sparsity-exploiting method. In this paper, exact reconstruction sampling condition for limited-view problem is studied by verifying the uniqueness of solution in TV minimization model. Uniqueness is tested by solving a convex optimization problem derived from the sufficient and necessary condition of solution uniqueness. Through this method, the sufficient sampling number of exact reconstruction is quantified for any fixed phantom and settled geometrical parameter in the limited-view problem. This paper provides a reference to quantify the sampling condition. Using Shepp-Logan phantom as an example, the experiment results show the quantified sampling number and indicate that an object would be accurately reconstructed as the scanning range becomes narrower by increasing sampling number. The increased samplings compensate for the deficiency of the projection angle. However, a lower bound of the scanning range is presented, in which an exact reconstruction cannot be obtained once the projection angle is narrowed to this extent no matter how to increase sampling.

---

[*] Corresponding author: Linyuan Wang, National Digital Switching System Engineering and Technological Research Center, Zhengzhou 450002, Henan, China, E-mail: wanglinyuanwly@163.com.

# 1. Introduction

In X-ray computed tomography (CT), an object is reconstructed from projections obtained by measuring the attenuation of X-rays in different views. When projection data satisfy the Tuy–Smith condition [1, 2], an object can be reconstructed accurately. However, this condition cannot be usually satisfied in practice because of restrictions on various features, such as object size, scanning geometry, and radiation dose. Projections are generally only allowed to be collected from a limited view with a scanning range of less than 180°. Challenges in accurate reconstruction often arise when complete projections cannot be obtained because of a deficient scanning range. Classical image reconstruction methods [3–7] are unsuitable for such a case. Image reconstruction from limited-view can be converted into the inversion of an ill-posed matrix, and proper limitation condition and regularization will help to improve the quality of limited-view reconstruction [8].

Recently, iterative image reconstruction (IIR) algorithms have been developed for X-ray tomography [9–13] based on compressive sensing (CS) [14, 15]. These algorithms achieved accurate reconstruction by using less data than those required by classical methods. This capability is made possible by exploiting sparsity, i.e., few non-zeros in the image or of some transform applied to the image. In CT imaging, total variation (TV)-based CT image reconstruction has been proven to be experimentally capable of producing accurate reconstructions from incomplete data [16–19]. However, the theoretical results from CS extending to the CT setting are scarcely discussed. CS provides theoretical guarantees of accurate undersampled reconstruction for certain classes of random measurement matrices [20], but not for deterministic matrices, such as CT system matrices. Although the abovementioned reference empirically demonstrated that CS-based methods are allowed for undersampled CT reconstruction, fundamental understanding of this deduction is lacking, and the conditions in which this method may be confidently used are undetermined. One problem in uncritically applying sparsity-exploiting methods to CT is that no quantitative notion of sufficient sampling for exact reconstruction is known. Pan et al. (2009) attempted to explain what CS means for image reconstruction, and their work bridged the link from CS to CT [21]. The concept of Hadamard full sampling was proposed in [22] in 2011, and a method to evaluate the sufficient data when applying CS theory to CT reconstruction was proposed. Four sufficient-sampling conditions were presented in [23] in 2012 and established a reference point to evaluate sampling reduction. Jørgensen et al. (2012) denoted a sufficient view number in [24], which is useful for quantifying the relationship between image sparsity and the minimum number of views for accurate reconstruction. Wang et al. (2014) presented the necessary condition related to the sufficient number of samplings to develop the theory of quantitative relationship between image sparsity and minimum sampling number [25]. Jørgensen et al. (2014) empirically demonstrated sharp average-case phase transitions from no recovery to exact recovery across a range of image classes and sparse reconstruction methods based on sufficient and necessary condition of solution uniqueness [26].

These aforementioned papers mainly focused on sparse-view reconstruction. However, few theoretical or practical investigations are known to quantify sampling condition particularly for limited-view exact reconstruction. Jørgensen et al. (2013) demonstrated empirically a relation between sparsity of the image to be reconstructed and the average number of fan-beam views required for accurate reconstruction on 90° limited angular data [27], which firstly explored the analysis of sampling condition for accurate reconstruction in limited view problem, and could be used for understanding what undersampling levels to expect when reconstructing sparse images. However, the proposed method is empirically and there is no strict mathematical verification for sampling condition. Mathematical analysis on the scanning range at which exact reconstruction can be realized and how many samplings are needed in that range has not yet been proposed [21]. The existing sufficient or necessary conditions for exact reconstruction have not deepened insight into the mathematical or physical essence of the reconstruction model. A precise mathematical study on sampling condition would help to develop a more efficient algorithm for the limited-view problem, which is inevitable to the issue of reducing radiation dose and of restricted scanning geometry [28–29]. Furthermore, the question of the suitable lower bound of the scanning range for limited-view exact reconstruction has not been answered until now. This issue needs to be studied urgently to evaluate object recoverability. In this paper, the sufficient and necessary condition of solution uniqueness is used to study the sampling number of exact reconstruction in the TV minimization model. This study attempts to quantify the sufficient sampling and to explore the lower bound of the scanning range for limited-view accurate reconstruction. In Section 2, a background of TV regularization is given and a method is proposed to quantify sufficient sampling for limited-view accurate reconstruction. Section 3 shows the experiment results and present a lower bound of scanning range. Section 4 elaborates the conclusions of this paper.

## 2. Methodology

### 2.1 Imaging model and system matrices

For most IIR algorithms, the data model is assumed to be linear and discrete-to-discrete (DD) [30–33]. We consider the inverse problem of recovering a signal $x \in \mathbb{R}^N$ from measurements $b \in \mathbb{R}^M$ as

$$b = Ax, \qquad (1)$$

where the elements in vectors $b$ and $x$ of finite dimensions denote data measurement and image-voxel values, and $A$ is the system matrix. To obtain the individual matrix elements, the line-intersection method is employed [34–36], where $A$ describes the intersection length of $i$ th ray with $j$ th pixel, and the dimensions of $A$ are $m = N_{view} \times N_{bins}$ rows (number of ray integrations) and $n = N_{pix}$ columns (number of variable pixels). The free parameter of this class is $N_{views}$, and $N_{pix} = N^2$, $N_{bins} = 2N$ are set. This description specifies the system matrix class for the present circular fan-beam CT study.

### 2.2 Sparse reconstruction algorithm in CT

Recently, for objects reconstruction from incomplete data, the issue that the system matrix has fewer rows than columns is a hotspot. System matrix $A$ has a non-trivial null space and the reconstruction problem is ill posed. The infinite number of the solution could be decreased by selecting the sparsest one, i.e., the one that has the fewest non-zeros, either in the image itself or after some transform has been applied to it. The reconstruction can be mathematically written as the solution of the constraint optimization problem

$$x^* = \arg\min_{x} \|\Psi(x)\|_0 \quad s.t. \ b = Ax, \tag{2}$$

where $\Psi$ is a sparsifying transform. $\|\ \|_0$ is the $l_0$-norm that denotes the number of non-zero elements, which measures the sparsity of its vector.

Computing the $l_0$-norm is an NP-hard problem, and other important results in CS involve relaxation of the non-convex $l_0$-norm to the convex $l_1$-norm

$$x^* = \arg\min_{x} \|\Psi(x)\|_1 \quad s.t. \ b = Ax, \tag{3}$$

where $\|\ \|_1$ describes the sum of absolute value which is easy to determine.

As suggested in [37], a potentially useful $\Psi$ would be to obtain sparse transform by computing the discrete gradient magnitude, which is zero within constant regions and non-zero along edges in CT. The $l_1$-norm applied to the gradient magnitude image is known as the anisotropic TV norm

$$\|x\|_{TV} = \|\Psi(x)\|_1 = \sum_{i} \|D_i x\|_2, \tag{4}$$

where $D_i$ computes the discrete gradient of pixel $i$. Because of $x$ is a vector, $l_2$-norm could be instead by $l_1$-norm. For anisotropic TV, we can rewrite the TV term

$$\|x\|_{TV} = \sum_{i} \|D_i x\|_1 = \sum_{j} \|D_j x\|_1, \tag{5}$$

where $D_j$ denotes the differential operator along direction $j$. In this paper, $D_1$ and $D_2$ denote the horizontal and vertical differential operator respectively for the two-dimensional form. Finally, the reconstruction problem can be formulated as

$$x^* = \arg\min_{x} \sum_{i} \|D_i x\|_1 \quad s.t. \ b = Ax. \tag{6}$$

**2.3 Exact reconstruction sampling condition based on solution uniqueness**

In many case, Eq. (5) needs to have unique solution. When there are more than one solution, the set of solutions is a convex set including an infinite number of solutions. In compressive sensing, having non-unique solution means that the object cannot be accurately recovered from the given data. In [38], the solution to various convex $l_1$-norm minimization problems would be unique if and only if a common set of conditions are satisfied. Moreover, the following conditions on system matrix $A$ are both sufficient and necessary to the uniqueness of solution $x*$.

**Condition 1** [38]: Under the definitions $I := \text{supp}(x^*) \subseteq \{1,2,...,n\}$ and $s := sign(x_I^*)$, the system matrix $A \in \mathbb{R}^{m \times n}$ has the following properties:

1. Submatrix $A_I$ has full column rank;

2. $\omega \in \mathbb{R}^m$ exists, obeying $A_I^T \omega = s$ and $\left\| A_{I^c}^T \omega \right\| < 1$.

The condition is extended to TV minimization problems and let $A \in \mathbb{R}^{m \times n}$ with $m < n$, $D \in \mathbb{R}^{n \times N}$ and $x^* \in \mathbb{R}^n$ with $I := \text{supp}(D^T x^*)$. The condition can be converted into the following theorem:

**Theorem 1** [26]**:** The solution $x^*$ is unique for the following optimization problem:

$$x^* = \arg\min_x \sum_j \left\| D_j x \right\|_1 \quad s.t.\ b = Ax. \tag{7}$$

if and only if

$$\ker(A) \cap \ker(D_{I^c}^T) = \{0\}, \tag{8}$$

and $\omega \in \mathbb{R}^m$ and $v \in \mathbb{R}^N$ such that

$$Dv = A^T \omega,\ v_I = \text{sign}(D_I^T x^*),\ \left\| v_{I^c} \right\|_\infty < 1. \tag{9}$$

The condition of zero-intersection of $\ker(A)$ and $\ker(D_{I^c}^T)$ can be checked numerically by evaluating whether the matrix $(A; D_{I^c}^T)$ has full rank, where semicolon means vertical concatenation. The second condition can also be tested by splitting the infinity norm into a two-sided inequality constraint and by solving the following optimization problem:

$$\begin{aligned}
&\min_{t, v, \omega}\ t \\
&\text{subject to}\quad -t\mathbf{1} \leq v_{I^c} \leq t\mathbf{1} \\
&\qquad\qquad A_I^T \omega = D_I v_I + D_{I^c} v_{I^c} \\
&\qquad\qquad v_I = \text{sign}(D_I^T x^*),
\end{aligned} \tag{10}$$

If the optimal $t^*$ is smaller than $1 - \varepsilon$ for $\varepsilon = 10^{-5}$, then the condition that inequality is satisfied strictly, and the solution of reconstruction model is unique.

For a DD reconstruction model, the solution is unique when its system matrix satisfies the sufficient and necessary condition. The condition that matrix $(A; D_{I^c}^T)$ has full rank is easy to verify.

This condition is usually satisfied for the CT system matrix in the TV minimization model, thus this paper mainly endeavors in computing $t$ to compare it with 1. The sampling number determines the system matrix dimension. For a certain system matrix, if solution uniqueness for the above optimization problem is satisfied under a certain sampling condition, then the solution is the exact reconstruction of this model. Therefore, solution uniqueness is equivalent to exact reconstruction in CT, and the object would be exactly reconstructed when the model solution is unique. With the use of the sufficient and necessary condition of solution uniqueness, the sampling condition for exact reconstruction in different scanning angles will be quantified in this study. Experiments will be conducted and presented in the following section to study the issue of sufficient sampling for limited-view exact reconstruction in CT.

## 3. Experiments and Discussion

We consider a digital 128×128 Shepp–Logan phantom as an example to verify the condition proposed above and to study the relationship between exact reconstruction and sampling number in limited-view. The size of system matrix $A$ are $m = N_{view} \times N_{bins}$ rows and $n = N_{pix}$ columns, where $N_{bins} = 256$ and $N_{pix} = 16384$. The source-to-object distance and the source-to-detector distance are 300mm and 800mm, respectively. The pixel of detector is 0.148mm and the voxel of reconstruction is $0.111*0.111 mm^2$. CVX, which is widely adopted to solve various convex problems, is used to compute $t$ in the above theorem.

### 3.1 Limited-view reconstruction studies

We consider a 2-D fan-beam scanner configuration with $N_{view}$ projections equi-distributed over angle range. Reconstruction from limited-view utilizes projections collected within 180°. And we select different scanning angle ranges, including 150°, 120°, 90°, 60°, 40°, 30°, 20°, and 15°, to study the limited-view sampling condition. We also generate system matrix $A$ for different numbers of views, $N_{view} \in [2, 30]$. To vary the number of views, $t_i$ $(i=2,...,30)$ is computed by CVX, and the index $i$ represents the view number in this experiment. The minimum sampling number for exact reconstruction is the one that is mostly close to 1 but never beyond it. The minimum sampling numbers for limited-view exact reconstruction in different scanning ranges are shown in Table 1. The variation trend of $t$ for different system matrices is shown in Fig. 1.

Table 1 Minimum sampling number for limited-view exact reconstruction

| Scanning range | 150° | 120° | 90° | 60° | 40° | 30° | 20° | 15° |
|---|---|---|---|---|---|---|---|---|
| Minimum sampling | 8 | 8 | 9 | 12 | 12 | 13 | 14 | 16 |
| $t$ | 0.9202 | 0.9869 | 0.8870 | 0.9436 | 0.9915 | 0.9477 | 0.9511 | 0.8880 |

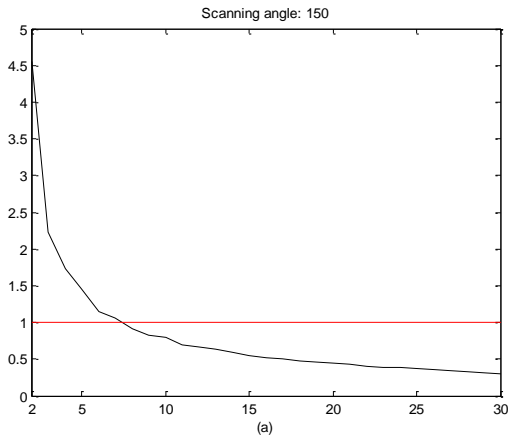
(a)

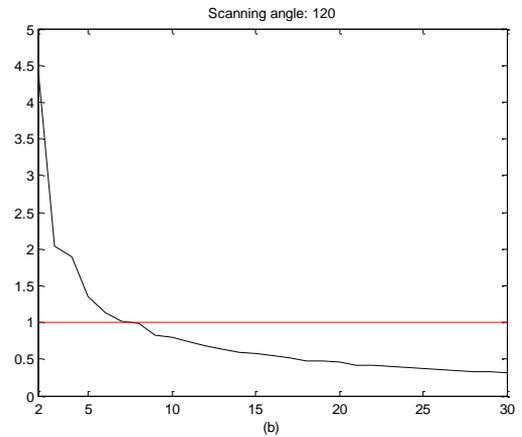
(b)

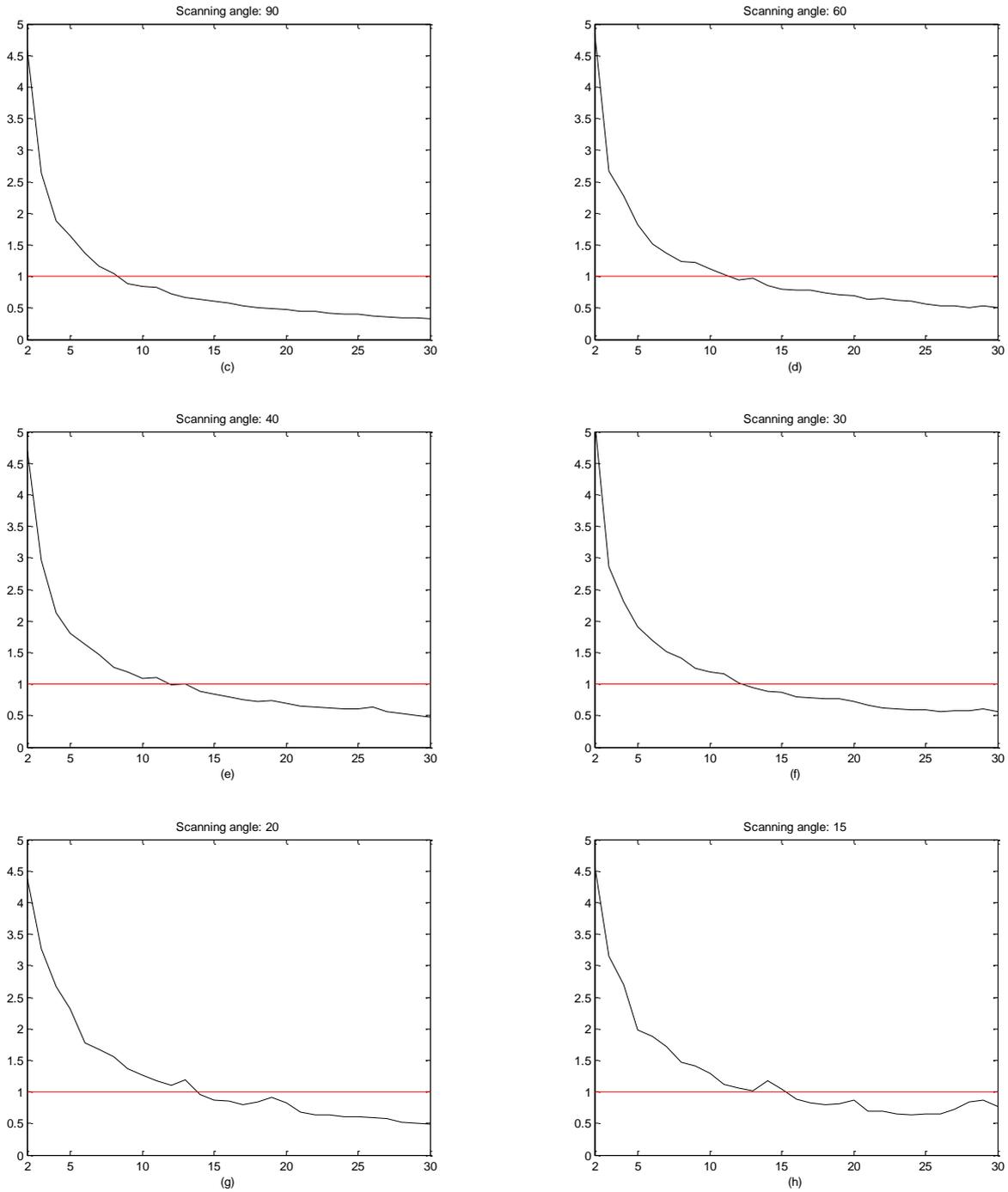

Fig. 1. Variation trend of $t$ for different system matrices. The x-label represents the sampling number which is [2, 30], and the y-label represents the value of $t$ which is [0, 5]. The red line indicates the threshold of exact recovery. Exact reconstruction occurs when $t$ is below the red line, not otherwise. The curve of (a)–(h) represents the scanning ranges 150°, 120°, 90°, 60°, 40°, 30°, 20°, and 15°.

As shown in Table 1 and Fig. 1, the sufficient sampling number for exact reconstruction increases gradually when the scanning range becomes narrow. Although the sufficient sampling number is the same in some scanning ranges, such as 150° and 120° and 60° and 40°, the value of $t$ for the same sampling number is different in these scanning ranges, and a smaller range computes larger $t$. The sampling number of exact reconstruction shows an increasing trend in general. Consequently, the result indicates that an object can be reconstructed exactly as the scanning range becomes narrow by increasing the sampling number. The increased samplings help to compensate for the deficiency of the scanning range. We then establish a logical question surrounding the lower bound of the scanning range, in which an exact reconstruction from limited-view cannot be obtained despite increasing the number of samplings. We attempt to answer this question through the following experiment.

**3.2 Narrower-view simulation studies and the lower bound of scanning range**

To explore the lower bound, simulation experiments are performed from narrower scanning ranges, including 14°, 13°, 12°, 11°, and 10°. For the TV minimization model, given that the sampling number is restricted by the size of system matrix $A \in \mathbb{R}^{m \times n}$, which has more columns than rows, the maximum sampling number is 63 for a $128 \times 128$ phantom in this experiment. We generate system matrix $A$ for different numbers of views, $N_{view} \in [2, 40]$. The exact reconstruction sampling number for narrower ranges when $t < 1$ is shown in Table 2. The variation trend of $t$ for different system matrices from narrower scanning ranges is shown in Fig. 2.

Table 2 Sampling number of limited-view exact reconstruction from narrower ranges when $t < 1$.

| Limited-view | 14° | 13° | 12° | 11° | 10° |
|---|---|---|---|---|---|
| Sampling number | 15, 17–25, $\geq 27$ | 15, 16, 18–20, 23, $\geq 27$ | 18, $\geq 25$ | $\geq 23$ | None |

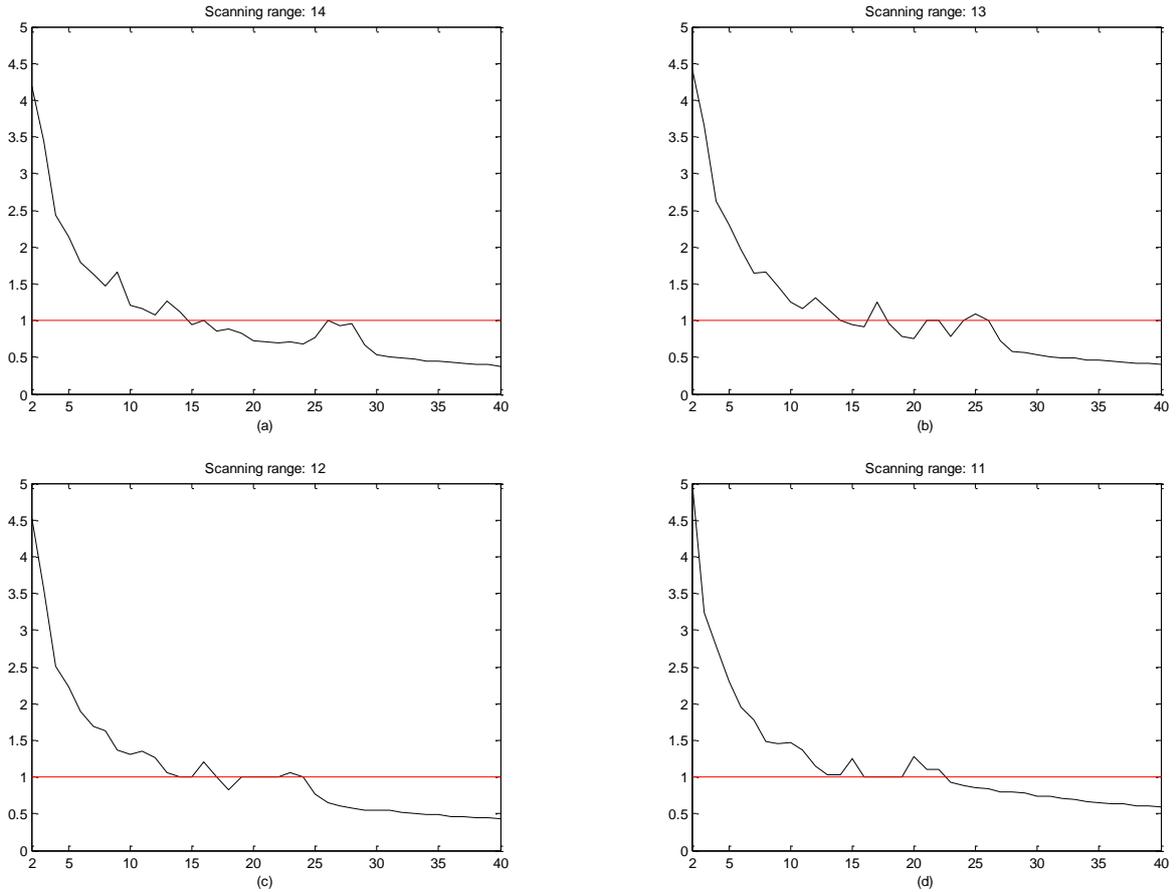

Fig. 2. Variation trend of $t$ for different system matrices from narrower scanning ranges. The x-label represents the sampling number, which is [2, 40], and the y-label represents the value of $t$, which is [0, 5]. The red line indicates the threshold of exact recovery. Exact reconstruction occurs when $t$ is below the red line, not otherwise. The curve of (a)–(d) represents the scanning ranges 14°, 13°, 12°, and 11°.

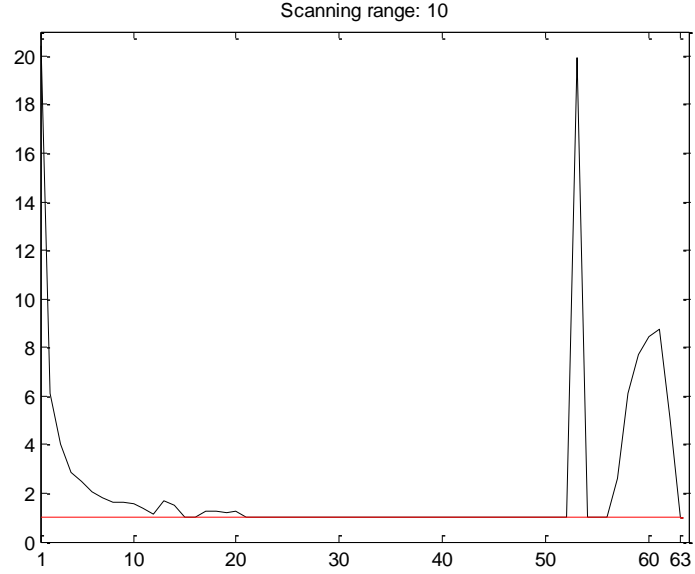

Fig. 3. Variation trend of $t$ for different system matrices from the scanning range 10°. The x-label represents the sampling number which is [1, 63], and the y-label represents the value of $t$ which is [0, 21]. The red line indicates the threshold of exact recovery. Exact reconstruction occurs when $t$ is below the red line, not otherwise.

Table 3 Values of $t$ for unusual sampling numbers when the scanning range is 10°

| Limited-view | 53 | 57 | 58 | 59 | 60 | 61 | 62 |
|---|---|---|---|---|---|---|---|
| $t$ | 19.9050 | 2.5921 | 6.1326 | 7.6862 | 8.4322 | 8.7694 | 5.1187 |

The results of Table 2 and Fig. 2 for the scanning ranges 14°, 13°, 12°, and 11° indicate that the sufficient sampling number of exact reconstruction from limited-view becomes unstable. The values of $t$ for four scanning ranges fluctuate when the sampling number varies from 5 to 30, and these values move up and down based on 1. Eventually, these values tend to exhibit a monotonic decrease and are strictly smaller than 1. This result indicates that the solution of reconstruction model is unique. Therefore, when the sampling number is large enough for scanning ranges 14°, 13°, 12°, and 11°, the exact reconstruction is able to be obtained always.

However, when the scanning range is narrowed to 10°, all the values of $t$ computed for different views are equal or greater than 1, as reflected by the variation trend of $t$ shown in Fig. 3. This result reveals that exact reconstruction cannot be obtained for all different samplings. A phenomenon is observed that the values of $t$ are significantly close to 1 when the sampling number is more than 20, except for 7 sampling points: 53, 57, 58, 59, 60, 61, and 62. A fluctuating interval is also observed, in which $t$ becomes unstable above the red line. As to the unusual phenomenon discussed above, one reason is certain that the computation of CVX was not completed before it goes up to the maximum iteration, and a guess is also possible that when the scanning range narrows to some extent, the recoverability is not only determined by the sampling number, but also the structure of system matrix. Further theoretical work is needed to explain these issues. Even so, one conclusion is certain: when the scanning range narrows to 10°, the exact reconstruction cannot be obtained despite increasing the number of samplings. Therefore, 10° is the lower bound of the scanning range for limited-view reconstruction that the phantom cannot be exactly recovered from this projection angle.

## 4. Conclusions

A quantitative notion of exact reconstruction sampling condition is necessary to provide a reference to evaluate recoverability in the limited-view problem. On the basis of a sufficient and necessary condition of solution uniqueness for the TV minimization model, a method that can quantify the sampling number of the limited-view exact reconstruction was proposed for a fixed phantom and settled geometrical parameter in this paper. The experiment results showed the quantified sampling number and indicated that as the scanning range becomes narrow, the phantom would be reconstructed accurately by increasing samplings, which compensate for the deficiency of the projection angle. A lower bound of the scanning range was presented for exact reconstruction of the fixed phantom, in which an exact reconstruction cannot be obtained once the scanning range narrows to this lower bound regardless of whether the number of sampling is increased. The proposed method provides a reference to evaluate the performance of reconstruction algorithms and to assess the difficulty of phantom reconstruction under different conditions. In the future study, it is necessary to improve the verification method of solution uniqueness to apply it into larger-scale reconstruction. The questions surrounding whether the sufficient projection number and the lower bound change along with the phantom sparsity should also be studied, this would promote further systematic study to describe the mathematics mechanism for limited-view problem.

## Acknowledgements

This work supported by the National High Technology Research and Development Program of China (863 Subject No. 2012AA011603), and the National Natural Science Foundation of China (No. 61372172).

## Reference

[1] H. Tuy. An inversion formula for cone-beam reconstruction. SIAM J Apply Math 1983; 43: 546-52.
[2] B. Smith. Image Reconstruction from Cone-Beam Projections: Necessary and Sufficient Conditions and Reconstruction Methods. IEEE Trans Imag Proc 1985; 4: 14-25.
[3] F. Noo, R. Clackdoyle, and J. Pack. A two-step Hilbert transform method for 2D image reconstruction. Phys Med Bio 2004; 49: 3903-23.
[4] M. Defrise, F. Noo, R. Clackdoyle, et al. Truncated Hilbert transform and image reconstruction from limited tomography data. Inverse Prob 2006; 22: 1037-53.
[5] Y. Ye, H. Yu, and G. Wang. Exact interior reconstruction with cone-beam CT. Int J Biomed Imag 2008; 2007: 10693.
[6] H. Yu, Y. Ye, and G. Wang. Interior reconstruction using the truncated Hilbert transform via singular value decomposition. J X-ray Sci Technol 2008; 16: 243-51.
[7] Q. Xu, H. Yu, X. Mou, et al. Low-dose X-ray CT reconstruction via dictionary learning. IEEE

Trans Imag Proc 2012; 31: 1682-97.

[8] H. Gao, L. Zhang, Z. Chen, et al. Reviews of Image Reconstruction from Limited-angle. CT Theory App 2006; 15: 46-50.

[9] J. Song, Q. Liu, G. A. Johnson, et al. Sparseness prior based iterative image reconstruction for retrospectively gated cardiac micro-CT. Med Phy 2007; 34: 4476-83.

[10] G. Chen, J. Tang, and S. Leng. Prior image constrained compressed sensing (PICCS): A method to accurately reconstruct dynamic CT images from highly undersampled projection data sets. Med Phys 2008; 35: 660-3.

[11] E. Sidky, X. Pan, I. S. Reiser, et al. Enhanced imaging of microcalcifications in digital breast tomosynthesis through improved image-reconstruction algorithms. Med Phys 2009; 36: 4920-32.

[12] J. Bian, J. Siewerdsen, X. Han, et al. Evaluation of sparse-view reconstruction from flat-panel-detector cone-beam CT. Phys Med Bio 2010; 55: 6575-99.

[13] L. Ritschl, F. Bergner, C. Fleischmann, et al. Improved total variation-based CT image reconstruction applied to clinical data. Phys Med Bio 2011; 56: 1545-61.

[14] D. Donoho. Compressed sensing. IEEE Trans Inf Theory 2006; 52: 1289-306.

[15] E. Candes, J. Romberg, and T. Tao. Stable signal recovery from incomplete and inaccurate measurements. Comm Pure Appl Math 2006; 59: 1207-23.

[16] M. Defrise, C. Vanhove, and X. Liu. An algorithm for total variation regularization in high-dimensional linear problems. Inverse Prob 2011; 27: 257-303.

[17] X. Han, J. Bian, D. Eaker, et al. Algorithm-enabled low-dose micro-CT imaging. IEEE Trans Med Imag 2011; 30: 606-20.

[18] E. Sidky, C. Kao, and X. Pan. Accurate image reconstruction from few-views and limited-angle data in divergent-beam CT. J X-Ray Sci Technol 2006; 14: 119-39.

[19] E. Sidky and X. Pan. Image reconstruction in circular cone-beam computed tomography by constrained, total-variation minimization. Phys Med Bio 2008; 53: 4777-807.

[20] E. J. Candes and M. B. Wakin. An introduction to compressive sampling. IEEE Signal Process Mag 2008; 25: 21-30.

[21] X. Pan, E. Sidky and M. Vannier. Why do commercial CT scanners still employ traditional filtered back-projection for image reconstruction? Inverse Prob 2009; 25: 123009.

[22] J. Jørgensen, E. Sidky and X. Pan. Analysis of discrete-to-discrete imaging models for iterative tomographic image reconstruction and compressive sensing. 2011; arXiv: 1109.0629.

[23] J. Jørgensen, E. Sidky and X. Pan. Quantifying admissible undersampling for sparsity-exploiting iterative image reconstruction in X-ray CT. IEEE Trans Med Imag 2013; 32: 460-73.

[24] J. Jørgensen, E. Sidky, P. Hansen, et al. Quantitative study of undersampled recoverability for spasrse image in computed tomography. 2012; arXiv: 1211.5658.

[25] L. Wang, H. Zhang, A. Cai, et al. System matrix analysis for sparse-view iterative image reconstruction in X-ray CT. J X-ray Sci Technol 2015; 23: 1-10.

[26] J. Jørgensen, C. Kruschel, and D. Lorenz. Inverse. Testable uniqueness conditions for empirical assessment of undersampling levels in total variation-regularized x-ray CT. Prob Sci En 2014; 23: 1283-305.

[27] J. Jørgensen, E. Sidky, and X. Pan. Connecting image sparsity and sampling in iterative reconstruction for limited angle X-ray CT. Proc. 12th Int. Meet. on Fully 3-D Image Reconstruction in Radiology and Nuclear Medicine. California: Lake Tahoe 2013; 169-72.

[28] L. Zeng, J. Guo, and B. Liu. Limited-angle cone-beam computed tomography image reconstruction by total variation minimization and piecewise-constant modification. J Inverse

Ill-Pose P 2013; 21: 735-54.

[29] Z. Bian, J. Ma, J. Huang, et al. SR-NLM: A sinogram restoration induced non-local means image filtering for low-dose computed tomography. Comput Med Imag Grap 2013; 37: 293-303.

[30] H. Barrett and K. Myers. Foundations of Image Science. Hoboken: Wiley-Interscience Press; 2004.

[31] J. Chen, L. Li, L. Wang, et al. Fast parallel algorithm for three-dimensional distance-driven model in iterative computed tomography reconstruction. Chin Phys B 2015; 24: 028703.

[32] G. Gullberg and G. Zeng. A reconstruction algorithm using singular value decomposition of a discrete representation of the exponential radon transform using natural pixels. IEEE Trans Nuc Sci 1994; 41: 2812–19.

[33] R. Huesman. The effects of a finite number of projection angles and finite lateral sampling of projections on the propagation of statistical nerrors in transverse section reconstruction. Phys Med Bio 1977; 22: 511-21.

[34] D. Man and S. Basu. Distance-driven projection and backprojection in three dimensions. Phys Med Bio 2004; 49: 2463-75.

[35] R. Siddon. Fast calculation of the exact radiological path for a three-dimensional CT array. Med Phys 1985; 12: 252-5.

[36] P. Joseph. An Improved Algorithm for Reprojecting Rays through Pixel Images. IEEE Trans Med Imag 1982; 1: 192-6.

[37] E. Candes, J. Romberg, and T. Tao. Robust uncertainty principles: Exact signal reconstruction from highly incomplete frequency information. IEEE Trans Inf Theory 2006; 52: 489-509.

[38] H. Zhang, W. Yin, and L. Cheng. Necessary and sufficient conditions of solution uniqueness in L1 minimization. J Optimiz Theory App 2015; 164: 109-22.